\documentclass[aps,prl,showpacs,nobibnotes,twocolumn,preprintnumbers,
nobalancelastpage,superscriptaddress]{revtex4}

\usepackage{graphicx}
\usepackage{amsmath}
\usepackage{latexsym}
\usepackage{dcolumn}
\usepackage{bm}
\input{epsf}
\newcommand{\be}{\begin{equation}}
\newcommand{\ee}{\end{equation}}
\newcommand{\ba}{\begin{eqnarray}}
\newcommand{\ea}{\end{eqnarray}}
\newcommand{\bi}{\begin{itemize}}
\newcommand{\ei}{\end{itemize}}
\newcommand{\bfi}{\begin{figure}
\epsfxsize=9cm
\epsffile}
\newcommand{\efi}{\end{figure}}
\newcommand{\la}{\lesssim}
\newcommand{\ga}{\gtrsim}

\begin{document}
\preprint{FERMILAB-PUB-05-260-A}
\title{
Mapping dark matter with cosmic magnification
} 
\author{Pengjie Zhang}
\email{pjzhang@shao.ac.cn, pen@cita.utoronto.ca}
\affiliation{Center for particle astrophysics,
Fermi National Accelerator Laboratory, Batavia, IL 60510-0500}
\affiliation{Shanghai Astronomical Observatory, Chinese Academy of Science, Shanghai, China}
\author{Ue-Li Pen}
\affiliation{Canadian institute for theoretical astrophysics,
  University of Toronto, 60 St. George street, Toronto, ON M5S 3H8}
\begin{abstract}
We develop a new tool to generate statistically precise dark matter maps 
from the cosmic magnification of galaxies with distance estimates.  We
show how to overcome the intrinsic clustering problem using the slope of the
luminosity function, because magnificability changes strongly over the luminosity function, while  
intrinsic clustering only changes weakly.  This may allow precision
cosmology beyond most current systematic limitations.  
SKA is able to reconstruct projected matter density map at smoothing scale
$\sim 10^{'}$ with S/N$\geq 1$, at the  rate of  $200$-$4000$ deg$^2$
 per year, depending on the abundance and evolution of 21cm emitting
galaxies. This power of mapping dark matter is comparable to, or even
better than that of cosmic shear from deep optical surveys or 21cm surveys.   
\end{abstract}
\maketitle

{\it Introduction.---}
The precision mapping of the universe, and the accurate determination
of cosmological parameters have been enabled by the recent generation
of cosmic microwave background(CMB) experiments, galaxy and lensing
surveys, and new analysis 
techniques.  Weak gravitational lensing has emerged with a promising
future of mapping dark matter directly, which would allow the
inference of the state of the universe, including its dynamics and the
nature of dark energy.  Lensing is free from modeling assumptions, and
can be accurately predicted from first principles.  Several major
surveys are underway, under construction or in the planning
stage. Currently, most attention 
has focused on using the lensing induced {\it cosmic
  shear}\cite{Shear}. But such an approach is subject to a 
  series of   
difficult experimental systematics \cite{Systematics}.   {\it CMB
  lensing}\cite{CMBlensing1} and {\it  21cm
  background lensing} \cite{21cmlensing} are promising. But contaminations such
  as the  kinetic Sunyaev Zeldovich effect  \cite{CMBlensing2} and/or
  non-Gaussianity may degrade their accuracy. In this paper
we will address an  
alternative approach, the lensing induced {\it cosmic magnification},
which is not subject to the known problems, and could
provide a robust statistical signal.

Traditionally, intrinsic  
clustering had presented a serious problem to measurement of cosmic
magnification.  The observable quantity is the surface density of
galaxies above some flux threshold.  A variation in this surface
density is then interpreted as lensing.  Unfortunately, intrinsic
clustering is usually larger than the lensing induced signal.
By utilizing the redshift information, intrinsic
clustering can be effectively eliminated in lensing correlation
functions\cite{Scranton05,Zhang05}.   In this paper, we further show
that, beyond the above statistical lensing measurement,  2D
convergence $\kappa$ 
maps can be reconstructed  with lower systematics 
 and larger sky coverage than cosmic shear maps, by utilizing both 
the redshift and flux information of galaxies. 2D $\kappa$
maps not only  provide independent and robust constraints on
cosmology, but also are complementary to traditional shear maps.  It
allows one to explicitly and locally solve for non-reduced shear, an
independent mode of checking E-B decomposition, and break the
mass-sheet degeneracy\cite{Schneider}.

{\it Cosmic magnification.---}  Cosmic magnification causes
coherent changes in the apparent galaxy number density.  Let $N_{ij}$ be the  
observed number of galaxies (including false peaks) at the $i$-th flux 
bin and  $j$-th redshift bin,  falling into an angular pixel centered
at direction $\hat{n}$ 
with angular size $\theta$. It can be expressed as 
\ba
\label{eqn:cm}
N_{ij}(\hat{n})=\bar{N}_{ij}+\bar{N}^r_{ij}\left[W_{ij}\kappa_j(\hat{n})+\delta_{g,ij}(\hat{n})\right]+\delta
N_{P,ij}(\hat{n})\ .
\ea
The signal $W\kappa$ has unique
dependence on galaxy flux through  $W=2(\alpha-1)$. Here,  $\alpha=-d
\ln[dn/dF]/d\ln F-1$ and $dn/dF$ is the mean number of observed
galaxies per flux interval
\footnote{The observed $dn/dF$ is convolved with system 
noise. Because there are more dwarf galaxies than massive ones,
noise makes the observed $dn/dF$ both larger and
steeper, in the flux range that SKA can probe at $z\ga 2$. The overall effect is that system noise in
  flux measurements increases the cosmic magnification
signal and strengthens the result in this paper. For simplicity,
we neglect this complexity.}
$\bar{N}_{ij}= 
\bar{N}^r_{ij}+\bar{N}^f_{ij}$, $\bar{N}^r_{ij}$, $\bar{N}^f_{ij}$
are  the mean number of detections, real galaxies and false peaks,
respectively. $\delta_g$ and $\delta N_P$ are  galaxy  intrinsic
clustering and Poisson fluctuation, respectively.

Our goal is to recover $\kappa$ of each angular pixel, given
observables  $N_{ij}$, $\bar{N}^r_{ij}$, $W_{ij}$ and
$\bar{N}_{ij}$
\footnote{Cosmic magnification does not change the
  averaged galaxy   
spatial and flux distribution, up to $O(\kappa^2)\sim 10^{-4}$
accuracy.  The sky coverage of SKA is $\ga 100$
deg$^2$. Thus for each redshift and flux bin, there are $\ga 4000$
angular pixels with size $\theta \sim 10^{'}$ and $\ga 10^5$ galaxies
across the survey sky, so $\bar{N}_{ij}$ can be measured
accurately. $\bar{N}^f_{ij}$ can be accurately 
predicted, since system noise is Gaussian and the dispersion $S_{\rm
  sys}$ is specified for each survey. The number of false peaks with
flux above $n$-$\sigma$, or $nS_{\rm sys}$  per redshift interval per
  beam is 
$[1.4\ {\rm Ghz}/\Delta \nu
  (1+z)^2]{\rm  Erfc}[n/\sqrt{2}]/2$.  $\Delta \nu$ is chosen to be
  the frequency 
width corresponding to $100$ km/$s$ velocity dispersion at redshift
z\cite{Zhang05}. Thus, one can accurately predict $\bar{N}^r_{ij}$ and
$W_{ij}$.}. 
We consider SKA\footnote{SKA:http://www.skatelescope.org/}, which can detect
  $\sim 10^8$ high z galaxies through the neutral hydrogen 21cm
  emission line.  
$\kappa$ has typical value $\sim 1\%$. To beat down Poisson
fluctuations, $\ga 10^4$ galaxies per  
angular pixel are required.  Traditionally, objects are selected at a
5$\sigma$ cut, where one can neglect the fraction of false detections.  This of
course  also discards the majority of the signal.  
With a $0.5$-$\sigma$ cut, one can reduce
  Poisson noise at $\theta\sim  10^{'}$. To
  increase lensing signal while reducing $\delta_g$ contamination, we focus 
on source redshifts $z\ga 2$.  After averaging over the full redshift range
$z\geq 2$,   $\delta_g$ is still several times
larger than $\kappa$.  However,  $\delta N_P$ and $\delta_{g,ij}$ have
different flux dependence to that of the signal.  Weighting each galaxies
  by some  function of their flux can  suppress
  the prefactors of $\delta_g$ and $\delta N_P$. Intuitively, Eq. 1
  implies the optimal estimator to be linear in $N_{ij}$. 

The predictions rely on  the assumed HI mass
function $n(M_{\rm HI},z)$. We extrapolate the locally 
observed $n(M_{\rm HI},z)=n_0(z)(M_{\rm
  HI}/M_*)^{-1.2}\exp(-M_{\rm HI}/M_*)$\cite{Zwaan97} to high
redshifts either 
assuming no evolution in both $n_0$ and $M_*$ ({\it conservative case}) or
$n_0(z), M_*(z) \propto (1+z)^{1.45}\exp(-z/2.6)$ ({\it
  realistic case}), which is calibrated against Lyman-$\alpha$
observations (refer to \cite{Zhang05} for details).
We adopt a flat 
$\Lambda$CDM cosmology with $\Omega_m=0.3$, $\Omega_{\Lambda}=0.7$,
$h=0.7$, $\sigma_8=0.9$, the primordial power index $n=1$, BBKS
transfer function\cite{BBKS} and Peacock-Dodds fitting formula for the
  nonlinear density power spectrum \cite{Peacock96}.

{\it The optimal estimator.---}
Since $z\ga 2$ galaxies are mainly
lensed by matter at $z\la 1$, $\kappa=A-B/\chi(z)$ is an excellent
approximation, where $A$ and $B$ are two constants and $\chi$ is the
comoving angular diameter distance. Since $\chi(z)$
varies slowly at $z>2$, one can approximate
$\kappa(\chi,\hat{n})\simeq \langle \kappa\rangle=   
\kappa(\langle \chi\rangle,\hat{n})$, where $\langle \chi\rangle=\sum
\bar{N}^r_{ij}/\sum_{ij} 
\chi_j^{-1}\bar{N}^r_{ij}$ is the effective distance to lens
\footnote{The
approximation $\kappa\simeq 
\langle \kappa\rangle$  simplifies the derivation of the
optimal estimator significantly, though its accuracy   can be as bad as
$\sim \pm 20\%$, at each redshift bins. But after averaging over many
redshift bins,  corrections in different bins effectively cancel. For
the optimal estimators derived (Eq. 3 \& 4), one can derive the
unbiased expression of $\langle \kappa\rangle$ such that $\langle
\kappa\rangle =\langle \hat{\kappa}\rangle$.}. 
In the limit that $\bar{N}_{ij}\gg 1$, Poisson fluctuations become 
Gaussian. The likelihood function of $\kappa$ at an angular pixel,
marginalized over $p(\delta_{g,11}\cdots£¬\delta_{g,ij})$, the probability
distribution of $\delta_{g,ij}$ of this angular pixel, is   
\ba
\label{eqn:fulllikilihood}
L&\propto&\int
\exp\left[-\sum_{ij}\frac{[N_{ij}-\bar{N}_{ij}-\bar{N}^r_{ij}(W_{ij}\kappa+\delta_{g,ij})]^2}{2\bar{N}_{ij}}\right]\nonumber
\\
&&\times p(\delta_{g,11}\cdots,\delta_{g,ij})\prod_{ij}d\delta_{g,ij}\ .
\ea
We choose the redshift bin size $\Delta z\ga
0.2$  and angular pixel size  $\sim 10^{'}$ such that 
$\delta_{g,ij}$ of different redshift bins are uncorrelated. For this
choice,  the matter density dispersion of each redshift bin
$\sigma_{m}\la 0.1$.  This verifies the neglect of  high order term $\delta_g
\kappa$ in  Eq. 1 \& 2. 

Since $\sigma_{m}\la 0.1$ and galaxy bias $b_g$ is
unlikely bigger than several\cite{bias}, it is reasonable to assume
that galaxies are Gaussian distributed. Then  $p(\delta_{g,11}\cdots)$
is completely 
determined by the covariance matrix
$C_{i_1j_1;i_2j_2}\equiv \langle
\delta_{g,i_1j_1}(\hat{n})\delta_{g,i_2j_2}(\hat{n})\rangle$.  SKA can
directly and accurately measure the correlations of galaxy density
fluctuations between flux bins, which are the sum of $C_{i_1j_1;i_2j_2}$,
correlations induced by lensing and cross terms. In the interesting
range, $C_{i_1j_1i_2j_2}$ 
dominates. So one can take the measured sum as first guess of
$C_{i_1j_1;i_2j_2}$. Maximizing $L$, one 
obtains the optimal estimator  $\hat{\kappa}$ of 
$\kappa$. The reconstructed $\kappa$ can in turn be applied to
subtract  the lensing contribution in the covariance matrix
estimation. This can be done 
iteratively. Since the lensing contribution is small, such iteration
should be stable and converge quickly. 

The properties of high redshift 21cm emitting galaxies are currently
poorly known.  It is likely that they trace the underlying dark matter at
some level, and that galaxies of different luminosities are correlated
to each other.  We consider this case first, and then the extreme
stochastic biasing limit\cite{stoch} where galaxies of different flux are
uncorrelated with each other. These two cases correspond to  the worst and
best cases for the $\kappa$ reconstruction, respectively. 

{\it Deterministic biasing.---} We first consider the case that $\delta_{g,ij}$
of different flux bins (but of the same redshift bin) are linearly
correlated, namely, $\delta_{g,ij}=b_{ij}\delta_j$, where $\delta_j$
is the dark matter density of the $j$-th redshift bin. As discussed
above, $b_{ij}$ can be measured iteratively. Marginalizing over $\delta_j$, we obtain
\ba 
\label{eqn:lower}
L&\propto& \exp\left[-\frac{(\kappa-\hat{\kappa})^2}{2(\Delta
    \kappa)^2}\right]\ ,\nonumber\\
\hat{\kappa}&=&(\sum_j S_j-\frac{B_jQ_j}{A_j})(\Delta \kappa)^2\ ,\nonumber\\
\Delta \kappa&=&\left(\sum_j T_j-\frac{Q_j^2}{A_j}\right)^{-1/2}\\
&\leq &\left(\bar{N}\langle W^2\rangle-\bar{N}\frac{\langle
  Wb\rangle^2}{\langle b^2\rangle}\right)^{-1/2}\ .\nonumber
\ea
Here $A_j=\sum_i (\bar{N}^r_{ij}b_{ij})^2/\bar{N}_{ij}+1/\sigma_j^2$, 
$B_j=\sum_i (N_{ij}-\bar{N}_{ij})\bar{N}^r_{ij}b_{ij}/\bar{N}_{ij}$, $
Q_j=\sum_i \bar{N}^{r,2}_{ij}W_{ij}b_{ij}/\bar{N}_{ij}$, $S_j=\sum_i
(N_{ij}-\bar{N}_{ij})\bar{N}^r_{ij}W_{ij}/\bar{N}_{ij}$, and
$T_j=\sum_i(\bar{N}^r_{ij}W_{ij})^2/\bar{N}_{ij}$.  $\bar{N}$ is the mean
number of galaxies in each angular 
pixel. $\langle \cdots\rangle$ are weighted by galaxies with the
noise from false peaks taken into account. 
\begin{table}
\begin{tabular}{ccccc}
\hline\hline
               &conservative   &realistic \\
$n$-$\sigma$   &$n_g$, $\langle W\rangle$,$\langle W^2\rangle$&$n_g$, $\langle
               W\rangle$,$\langle W^2\rangle$ \\
0.5            &123,-0.44,1.1 &\ 419,-0.85, 1.5 \\
1.0            &76,\ -0.21,1.2  &\ 290,-0.70, 1.4 \\
2.0            &40,\ 0.11,1.6   &\ 184,-0.49,1.4 \\
5.0            &13,\ 0.98,3.7  &82,\ 0.1,\ 1.8 \\ 
10.0           &3.9,\ 2.0,  7.7&\ 36,\ 0.73 \  2.9 \\
\hline\hline
\end{tabular}
\caption{The predicted number of galaxies $n_g/1^{'2}$  at
  $z\geq 2$, $\langle W\rangle$ and $\langle W^2\rangle$ for SKA deep
  survey with integration time $18$ days/deg$^2$. \label{table:ska}} 
\end{table}

{\it Maximal stochasticity.---}
Stochasticity eases the subtraction of the intrinsic clustering signal.
In this case, $\delta_g$ of different bins are uncorrelated. We have
\ba
\label{eqn:upper}
L&\propto& \exp\left[-\sum_{ij}\frac{[N_{ij}-\bar{N}_{ij}-\bar{N}^r_{ij}W_{ij} 
    \kappa]^2}{2\sigma_{ij}^2} \right]\ ,\nonumber \\
\hat{\kappa}&=&\frac{\sum_{ij}(N_{ij}-\bar{N}_{ij})\bar{N}^r_{ij}W_{ij}/\sigma^2_{ij}}{\sum_{ij}[\bar{N}^r_{ij}W_{ij}]^2/\sigma^2_{ij}}\
,\nonumber \\
\Delta \kappa&=&[\partial^2\ln L/\partial\kappa^2]^{-1/2}=\left[\sum_{ij}
\frac{[\bar{N}^r_{ij}W_{ij}]^2}{\sigma^2_{ij}} \right]^{-1/2}\nonumber
\\
&\geq& [\bar{N}\langle W^2\rangle]^{-1/2}\  \ '='\ when\
\bar{N}^r_{ij}\sigma^2_{g,ij}\rightarrow 0\ .
\ea
Here, $\sigma^2_{ij}=\bar{N}_{ij}+\bar{N}^{r,2}_{ij}\sigma_{g,ij}^2$, where
the first term is the shot noise  and the second term is the intrinsic
fluctuation of galaxy number distribution.  The conditions
$\bar{N}^r_{ij}\sigma^2_{g,ij}\la 0.01\bar{N}_{ij} \rightarrow 0$ and
$\bar{N}_{ij}\gg 1$ (for Gaussianity) can both  be satisfied since
    galaxy bias $b_g$ is 
unlikely bigger than several\cite{bias}. A similar estimator has been
derived by \cite{Menard02}. In two estimators, $\langle W^2\rangle$
    and $\langle Wb\rangle$ are two key ingredients and reflect the
    key role of flux information.

\bfi{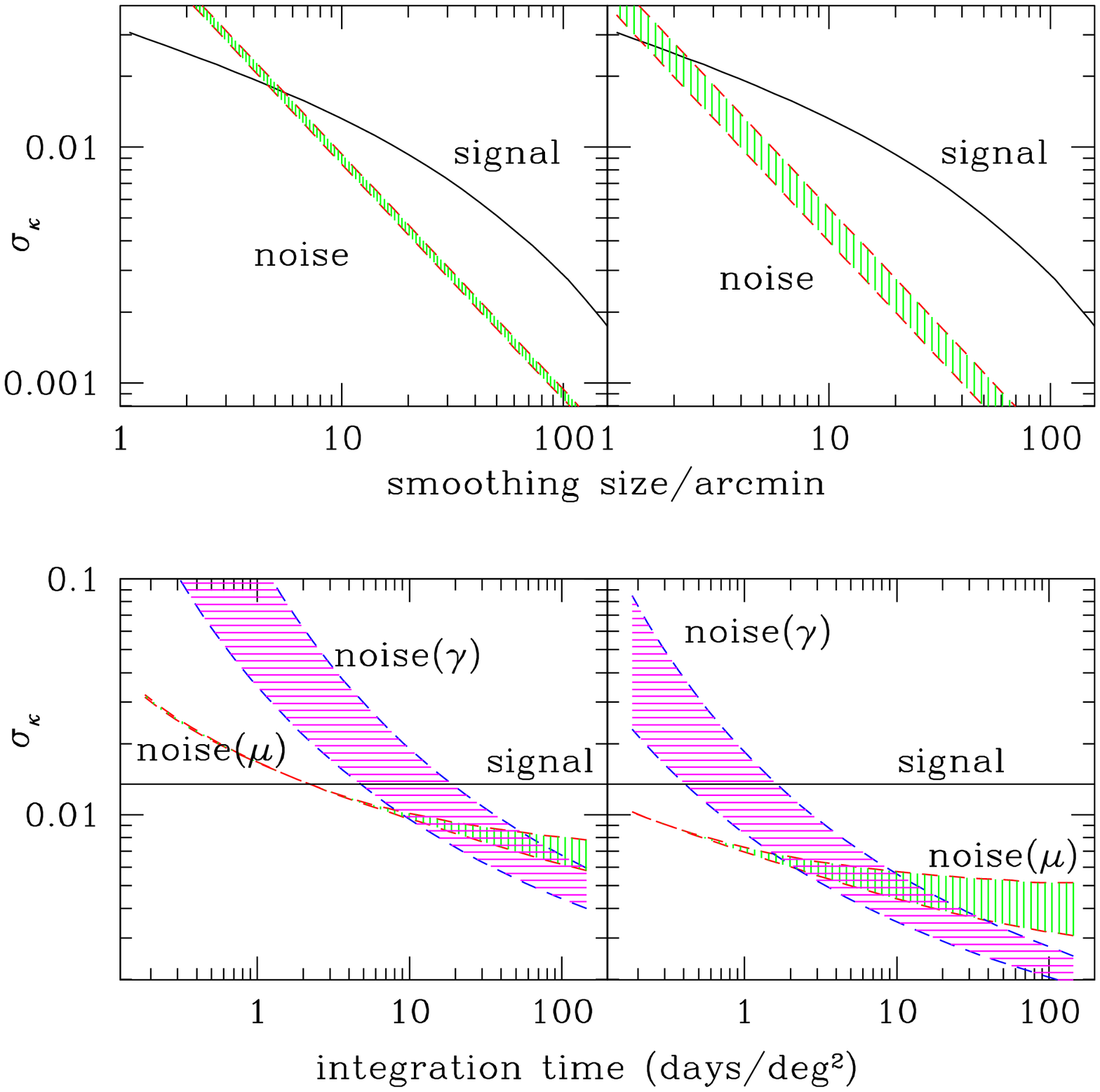}
\caption{Estimated power of SKA to recover the lensing convergence
  map. Left panels and right panels correspond to the conservative
  case and the  realistic case.  
We only use galaxies at $z>2$ above a
  $0.5$-$\sigma$ detection 
  threshold and assume unity galaxy bias. In top panels,
  we fix the integration rate as 
  18 days/deg$^2$ and plot rms of  signal (black lines) and noise
  (shaded regions) as function of
  smoothing size.  Upper and lower limits of statistical  errors are
  calculated  by Eq. \ref{eqn:lower} and \ref{eqn:upper},
  respectively.   Bottom panels show the comparison between cosmic
  magnification ($\mu$) and 
  cosmic shear ($\gamma$) at $10^{'}$ smoothing scale. The lower and
  upper limit  of shear measurement error are estimated using $5$-$\sigma$ and
  $10$-$\sigma$ cut, respectively and adopting mean ellipticity $\langle
  \epsilon^2\rangle^{1/2}=0.54$, as expected for disk galaxies. SKA is
  likely capable of  
  reconstructing the $\kappa$ map with $S/N>1$ at the rate of $\sim 4000$
  deg$^2$/year. Cosmic shear quickly loses power
  when decreasing the integration time per
  deg$^2$ while cosmic magnification is less affected. \label{fig:error}}
\efi

{\it Results.---}
SKA is able to detect $n_g\ga 100\ {\rm arcmin}^{-2}$
galaxies at  $z\ga 2$ (table I).  For an integration time $t_{\rm
  int}=18$ days/deg$^2$, a S/N$\ga 2$  can be achieved  at $\theta\sim
10^{'}$ (fig. 1).  Deep survey configuration detects more faint
galaxies, which have $W\rightarrow -2$, mimic a constant $b$ and thus
do not contribute to the signal, due to  the $\langle W^2\rangle-\langle
Wb\rangle^2/\langle b^2\rangle$  facotr in Eq. 3. An optimal survey configuration
should have $\langle Wb\rangle\rightarrow 0$, which can be achieved
at $t_{\rm int}\sim 0.2$-$1$ day/deg$^2$ (fig. \ref{fig:error}).
Since $n_g$ above $0.5$-$\sigma$ decreases much more slowly than $t_{\rm
  int}$ (for example, for the evolution model, decreasing  $t_{\rm int}$ from $180$ days/deg$^2$
to  $4$ hours/deg$^2$, $n_g$ only decreases by a factor of $9$), it is still
likely to achieve S/N$>1$ at $\theta\sim 10^{'}$ and scan rate of
$\sim 4000$ deg$^2$ per 
year (fig. 2).  This will produce more lensing information ($\propto$
S/N$\times f_{\rm sky}^{1/2}$) in a one year SKA survey than
SNAP\footnote{SNAP: http://snap.lbl.gov/} will 
produce, which will cover $1000$ deg$^2$ sky area with S/N$\sim 2$ at
smoothing scale $\theta\sim 10^{'}$.  

Since the SKA will have $\sim 0.3^{''}$ resolution at $z\sim 2$, it can resolve
galaxies and measure cosmic shear.  An
intrinsic advantage  of cosmic magnification measurement over cosmic
shear measurement is that it does not require galaxies to be
resolved. Thus, dwarf galaxies which are too small and too
faint for reliable shear measurement  still contribute to
magnification measurement.
Cosmic magnification exceeds
cosmic shear at integration rate $\la 0.2$-$10$ days/deg$^2$
(fig.\ref{fig:error}). We note
that this comparison is 
conservative.  We have neglected all systematics of shear
measurement. For magnification estimation, we only select galaxies above 
a $0.5$-$\sigma$ detection threshold, or HI mass above ${\rm
  several}\times 10^8 M_{\odot}h^{-2}$. There are numerous galaxies 
with HI mass $\sim 10^{7}  M_{\odot}h^{-2}$\cite{Zwaan97}, which can
in principle be used to improve the measurement. We do not explore its
potential in this paper since the luminosity
function at the faint end is unclear.

Several uncertainties could degrade the signal separation. (1) The
HI mass function, which is the dominant factor, as 
can be seen from table I and fig. 1. Here we further  draw the
attention on the  slope of the HI mass
function. For an extreme case that 
$\alpha\rightarrow 1$ and $W\rightarrow 0$ over a large flux range,
the signal disappears. This effect  can be straightforwardly
estimated through the 
$\langle W^2\rangle$ and $\langle W b\rangle$ terms in Eq. 3 \&
4, once the HI mass function is measured. Since HI mass function at
high $z$ is effectively unknown, we postpone the discussion in this
paper. (2) The galaxy bias. For the case of deterministic biasing, if
$b_g\propto W$, flux information is no longer useful for the
separation and our method effectively fails. But since $b_g>0$, as
long as the survey is deep enough to  probe the faint end of galaxies
where $W<0$, $b_g$ can not always  mimic $W$ and the separation is
always possible. (3) The galaxy distribution. When $b_g$ is bigger
than several or smoothing size is smaller than several arc-minutes,
$\delta_g$ is non-Gaussian. In this case, the estimators described
above are no longer optimal. Optimal estimators for non-Gaussian
galaxy distribution should be further investigated.

{\it Applications.---}The reconstructed $\kappa$ map can be applied to
measure many lensing statistics. For this purpose, reconstructed
$\kappa$ can be noisy because  these statistics  generally average
over many angular pixels and achieve high S/N.  Then
the optimal estimator derived in this paper can be  applied to  each
narrow redshift bins and allows the lensing tomography. 
(1){\it The probability density function
  $p(\kappa)$}. $p(\kappa)$ as a function of $\kappa$ 
and smoothing angular size $\theta$ can provide independent constraints
on cosmology. 
 Recently \cite{Zhangtj05} showed that
the Wiener filter reconstruction of $p(\kappa)$ from noisy
convergence map can go deep into regions where $|\kappa/\Delta\kappa|\ll
1$.  We thus expect that $p(\kappa)$ can be
recovered accurately from SKA.  (2) 
{\it Lensing power spectrum and bispectrum}. 
The reconstructed $\kappa$ map barely has S/N$\ga 5$, so it is consistent
to neglect $\la 10\%$ higher order terms:  $O(\kappa^2)$ terms and
$\delta_g\kappa$ term neglected  in Eq. 1 and
$\kappa(\chi)-\langle\kappa\rangle$.  
But these terms should be taken into account for precision measurement
of lensing power  spectrum and bispectrum, since their  statistical
errors can reach $\sim 1\%$ 
accuracy\cite{Zhang05}.  For the linear estimator we derived,
contributions of these terms to the power spectrum and 
bispectrum can be straightforwardly and robustly predicted. So, there
is no need to derive a more complicated nonlinear estimator. (3)
{\it Cluster finding and cluster density profile}. This is a promising approach to break the cluster mass sheet degeneracy. In the
reconstructed maps, massive clusters at $z\sim 0.2$ show as high peaks
with strength $\kappa\sim 0.1$ and size $\sim 10^{'}$ and can be
easily identified. These clusters are excellent objects to measure the
geometry of the universe by the technique of lensing cross-correlation
tomography \cite{Tomography}. Since S/N is so high, one can choose
smoothing size $\sim 1^{'}$ and  measure the projected cluster density
profile. Exerting a prior on cluster density profile, the  
reconstruction  can be further improved \cite{Dodelson04}. When
$\kappa\rightarrow 1$, the weak lensing 
condition breaks and Eq. \ref{eqn:cm} no longer holds. By
utilizing the exact magnification equation, one can develop new
estimator, in analogy to the reduced shear reconstruction
\cite{Penstrong}. We leave this topic for further study.

We summarize our results.  Cosmic magnification is statistically more
sensitive than cosmic shear because it is possible to use the large
number of galaxies detected at low statistical significance.  Intrinsic
clustering can be subtracted because (1) magnification depends strongly on the
shape of the luminosity function, which varies significantly, while
intrinsic clustering depends weakly on the intrinsic luminosity
itself and (2) they have different redshift dependence.  Cosmic
magnification shows promise as a complementary 
technique to map the statistically precise distribution of matter,
which is not subject to most of the systematics of cosmic shear.  We
have worked through the specific numbers for the SKA, but the general
formalism would also apply to optical spectroscopic or photometric
redshift surveys.

{\it Acknowledgments.---} We thank Scott Dodelson for
many helpful conversations and careful proofreading. We thank Albert
Stebbins and Martin White for helpful discussions.  P.J. Zhang was
supported by the DOE and the NASA 
grant NAG 5-10842 at Fermilab.

\end{document}